\documentclass[12pt]{iopart}
\usepackage{iopams} \usepackage{graphicx}
\begin{document}

\title{Gravastar energy conditions revisited}

\author{Dubravko Horvat and Sa\v{s}a Iliji\'c}

\address{Department of Physics,
  Faculty of Electrical Engineering and Computing,
  University of Zagreb, Unska 3, HR-10\,000 Zagreb, Croatia}

\ead{dubravko.horvat@fer.hr, sasa.ilijic@fer.hr}

\begin{abstract}
We consider the gravastar model where the vacuum phase transition
between the de~Sitter interior
and the Schwarzschild or Schwarzschild--de~Sitter exterior geometries
takes place at a single spherical $\delta$-shell.
We derive sharp analytic bounds on the surface compactness ($2m/r$)
that follow from the requirement
that the dominant energy condition (DEC) holds at the shell.
In the case of Schwarzschild exterior,
the highest surface compactness is achieved with the stiff shell
in the limit of vanishing (dark) energy density in the interior.
In the case of Schwarzschild--de~Sitter exterior,
in addition to the gravastar configurations
with the shell under surface pressure,
gravastar configurations with vanishing shell pressure (dust shells),
as well as configurations with the shell under surface tension,
are allowed by the DEC.
Respective bounds on the surface compactness are derived for all cases.
We also consider the speed of sound on the shell
as derived from the requirement that the shell
is stable against the radial perturbations.
The causality requirement (sound speed not exceeding that of light)
further restricts the space of allowed gravastar configurations.
\end{abstract}


\section{Introduction}

The gravastar model has been proposed by Mazur and Mottola (MM)
as a possible end point of the gravitational collapse
of a massive body \cite{MazMot01,MazMotPNAS}.
As opposed to the classical scenario of the gravitational collapse
into a Schwarzschild black hole,
the gravastar concept avoids the formation of the black hole horizon.
Technically, the gravastar is a global, static,
spherically symmetric solution to the Einstein equations.
In the interior there is a segment of the de~Sitter geometry
which is matched to the exterior Schwarzschild geometry
by means of a spherical shell of matter.
In the original MM model the quantum phase transition
takes place within the boundary layer/shell,
but the full quantum treatment is immediately replaced
by the mean field approximation in which (classical)
Einstein equations are valid.
The de~Sitter geometry in the interior of the gravastar
implies constant positive (dark) energy density $\rho$
accompanied by the isotropic negative pressure $p=-\rho$.
Such energy-momentum content of the spacetime is equivalent
to introducing the positive cosmological constant $\lambda=8\pi\rho$
into the Einstein equations,
and can be understood as the gravitational vacuum,
or gravitational condensate,
which motivated the coining of the term gravastar
(\emph{gra}vitational \emph{va}cuum \emph{star}).
In this original model of MM,
the phase transition layer consists of two spherical $\delta$-shells
with vanishing surface energy density
but with non-vanishing surface pressure,
between which there is a thin layer of stiff matter ($\rho=p>0$).
The transition layer is positioned arbitrarily close
to where the horizon of the Schwarzschild geometry is expected to form.
In this way the gravastar model allows for arbitrarily high
gravitational surface red-shifts and is, for an external observer,
indistinguishable from being a true black hole.

A simplified gravastar model
was constructed by Visser and Wiltshire \cite{VissGS1}
in order to study the stability properties.
The phase transition layer was replaced by a single spherical $\delta$-shell.
The issue of stability could not be fully settled
because the equation of state for the shell is not available.
However, it was shown that physically reasonable equations of state
could lead to gravastar configurations
that are stable against radial perturbations.
This simplified gravastar model also motivated
a number of authors to consider different
vacuum geometries in the interior and the exterior.
This includes the models with anti-de~Sitter interior
and the Schwarzschild--de~Sitter exterior geometries
and also the Reissner--Nordstr\"om geometries \cite{CarterStableGS},
the Born--Infeld phantom gravastar \cite{BilicGS},
and the non-linear electrodynamics gravastar \cite{LoboNLED}.
A different development of the gravastar idea went in the direction
of constructing models with continuous profiles for the energy density
and the anisotropic pressures \cite{VissGS2,DeBenedGS1}.
In these models the energy momentum tensor
in the centre of the spherical structure satisfies $\rho=-p$
and through the layer of anisotropic principal pressures (crust)
smoothly joins with the exterior vacuum spacetime.
Gravastars with shells of finite thickness
but without $\delta$-shells at the boundaries with the vacuum regions
have been considered in Ref.~\cite{ChirentiRezzolla07}
and it has been pointed out that quasi-normal modes of oscillations
might be a possible way of distinguishing gravastars from black holes.
An account of the observational constraints placed on the gravastar model
has been given in Ref.~\cite{BroderickNarayan07}.
The gravastar models are also closely related
to a broader subject of dark energy stars,
see eg.\ Ref.~\cite{LoboStableDES}.
In fact, the ideas of astrophysical bodies
with the de~Sitter core representing
the gravitationally collapsed matter at extremely high density
can be traced back to the work
of Sakharov \cite{Sakharov66} and Gliner \cite{Gliner66}
and have been further investigated by Dymnikova
in a series of papers starting with \cite{Dym92},
see eg.\ Refs~\cite{Dym02,Dym07} and references therein.

Given that the physical mechanism behind the proposed vacuum phase transition
within the gravastar shell is essentially unknown,
the energy conditions of GR \cite{HawkingEllis73}
provide the most widely accepted framework
of model-of-matter independent constraints
one can impose on the `exotic matter' comprising the shell.
It is therefore important to fully investigate
the constraints on the gravastar model
that are imposed by the energy conditions.
In this work we will consider
the $\delta$-shell gravastar model of Ref~\cite{VissGS1}
and we will derive the sharp analytic bounds on the surface compactness
that follow from the requirement
that the dominant energy condition (DEC) is satisfied at the shell.
The DEC will include the weak (WEC) and the null energy conditions (NEC),
but not the strong energy condition (SEC).
Since the SEC is violated from the outset
by the dark energy in the interior of the gravastar,
it does not appear natural to require
that it holds throughout the phase transition layer.

In Sec.~\ref{sec:model} we review the $\delta$-shell gravastar model 
and we introduce the configuration variables
that we use through the rest of the paper.
In Sec.~\ref{sec:ec} we discuss the energy conditions of GR
imposed on the energy-momentum tensor of the gravastar shell
and in Sec.~\ref{sec:allowed} we carry out the detailed analysis
of the gravastar configurations allowed by the DEC.
In Sec.~\ref{sec:eos} we discuss the properties of the
equation of state as derived from the assumed stability of the gravastar
with respect to radial perturbations.
Conclusions are given in Sec.~\ref{sec:conc}.

\section{The gravastar model \label{sec:model}}

The line element for the interior and the exterior geometries
can be written using the geometrized units $G=1=c$ and the coordinates
$x^{\alpha}=(t,r,\vartheta,\varphi)$
in the form
  \begin{equation}
  \rmd s^2 = g_{\alpha\beta} \, \rmd x^{\alpha} \rmd x^{\beta}
          = - k\,\big(1-\mu(r)\big) \, \rmd t^2
            + \frac1{1-\mu(r)} \, \rmd r^2
            + r^2 \, \rmd \Omega^2
  \label{eq:gends2}
  \end{equation}
where $\rmd \Omega^2=\rmd\vartheta^2+\sin^2\vartheta\,\rmd\varphi^2$
is the metric on the unit 2-sphere,
$k>0$ is the time coordinate scaling constant,
and $\mu(r)$ is the \emph{compactness function}
related to the quasi-local mass function $m(r)$ by
  \begin{equation}
  \mu(r) = \frac{2m(r)}{r}
         = \frac2r \int_0^r 4\pi\, \bar{r}^2 \, \rho(\bar{r}) \, \rmd \bar{r},
  \end{equation}
where $\rho(r)$ is the energy density.
In general, to avoid the formation of the event horizon
within the spherical body the compactness must be less than unity.

In the interior of the gravastar shell
we have the segment of the de~Sitter geometry
with the constant energy density
$\rho_{\mathrm{int}}=\lambda_{\mathrm{int}}/8\pi\ge0$
and the compactness is given by
  \begin{equation}
  \mu_{\mathrm{int}}(r) = \frac{8\pi}3 \, \rho_{\mathrm{int}} \, r^2
           = \frac{\lambda_{\mathrm{int}}}{3}\,r^2, \qquad r<a.
  \label{eq:muint}
  \end{equation}
At $r=(3/\lambda_{\mathrm{int}})^{1/2}$ the interior compactness reaches unity
which corresponds to the position of the de~Sitter (cosmological) horizon.
Therefore the surface of the gravastar
must be located at the radius $r=a<(3/\lambda_{\mathrm{int}})^{1/2}$.

In the exterior we have the Schwarzschild--de~Sitter geometry
with the constant energy density
$\rho_{\mathrm{ext}}=\lambda_{\mathrm{ext}}/8\pi\ge0$
and mass parameter $M>0$.
The exterior compactness is given by
  \begin{equation}
  \mu_{\mathrm{ext}}(r)
           = \frac{2M}r + \frac{8\pi}3 \, \rho_{\mathrm{ext}}\, r^2
           = \frac{2M}r + \frac{\lambda_{\mathrm{ext}}}{3}\,r^2, \qquad r>a.
  \label{eq:muext}
  \end{equation}
With $\lambda_{\mathrm{ext}}=0$, $M>0$,
this is clearly the Schwarzschild geometry,
while with $\lambda_{\mathrm{ext}}>0$, $M=0$,
this is the de~Sitter geometry.
In the general case we are considering,
$\lambda_{\mathrm{ext}}>0$, $M>0$,
the gravastar surface must be located
in the region where $\mu_{\mathrm{ext}}<1$.
For this region to exist
the condition $9M^2\lambda_{\mathrm{ext}}<1$ must hold,
whereupon the radial coordinates of the two horizons
of the Schwarzschild--de~Sitter geometry which bound the region
are given by the roots of $\mu_{\mathrm{ext}}(r)=1$ (a cubic).
Thus, if the gravastar surface is located at $r=a$
such that $\mu_{\mathrm{int}}(a)<1$ and $\mu_{\mathrm{ext}}(a)<1$,
no horizon forms within the gravastar geometry
up to the outer horizon of the Schwarzschild--de~Sitter geometry.

Since the original motivation behind the MM gravastar model
was to construct a highly compact object that could represent
a gravitationally collapsed body of positive gravitational mass,
one expects the (dark) energy density inside the gravastar
to be greater than that of the surrounding space,
i.e.\ $\rho_{\mathrm{int}}\ge\rho_{\mathrm{ext}}$,
or equivalently $\lambda_{\mathrm{int}}\ge\lambda_{\mathrm{ext}}$.
We refer to the above as the ``gravastar requirement''.
Note also that the constants
$\lambda_{\mathrm{int}}$ and $\lambda_{\mathrm{ext}}$
are introduced for notational convenience;
however, $\lambda_{\mathrm{ext}}$ can be viewed
as the cosmological constant $\Lambda=\lambda_{\mathrm{ext}}$
present in the Einstein equations
provided that in the exterior of the gravastar shell
we set $\rho_{\mathrm{ext}}\to\rho_{\mathrm{ext}}'=0$
and in the interior we set
$ \rho_{\mathrm{int}}\to\rho_{\mathrm{int}}'
= ( \lambda_{\mathrm{int}} - \lambda_{\mathrm{ext}} ) / 8 \pi$.
Thus, in the picture with the cosmological constant,
demanding that the energy density in the interior
of the gravastar shell is non-negative, $\rho_{\mathrm{int}}'\ge0$,
is equivalent to our ``gravastar requirement''.
Therefore, although it \emph{is} technically possible to construct
the gravastar-like solutions to the Einstein equations
with $\rho_{\mathrm{int}}<\rho_{\mathrm{ext}}$ and $M>0$,
we will not consider them in this analysis.

Through the rest of the paper we will describe the gravastar configurations
in terms of the configuration variables defined as follows:
  \begin{eqnarray}
  x = \mu_{\mathrm{int}}(a) = \lambda_{\mathrm{int}} a^2 / 3, \nonumber \\
  y = \mu_{\mathrm{ext}}(a)
    = 2M/a + \lambda_{\mathrm{ext}} a^2 / 3 = 2M/a + z, \label{eq:xyz} \\
  z = \lambda_{\mathrm{ext}} a^2 / 3. \nonumber
  \end{eqnarray}
$x$ and $y$ are the values of the compactness on the interior
and the exterior side of the shell, while $z$ is the contribution
to the compactness at the exterior side of the shell
due to the energy density in the exterior.

For the gravastar geometry we require
  \begin{equation}
  0 \le z \le x \le y < 1.
  \label{eq:zxy}
  \end{equation}
$z$ and $x$ are non-negative
since the energy densities in the interior and in the exterior
are non-negative by assumption.
The condition $z\le x$ reflects the above mentioned `gravastar requirement'.
The condition $y\ge z$ ensures that the mass of the gravastar, $M$,
is non-negative, while $y\ge x$, as will be shown in Sec.~\ref{sec:ec},
ensures that the energy density of the gravastar shell is non-negative.  

The gravastar shell is a timelike hypersurface of constant
radius $r=a$ along which the interior and the exterior geometries are joined.
The Israel's thin shell formalism \cite{Israel66}
(for textbook coverage see eg.~\cite{VisserBook,PoissonBook})
allows one to assign the energy-momentum distribution to the hypersurface,
as required to make the resulting spacetime
a solution of Einstein equations.
The first junction condition requires that the
induced metric on the hypersurface be the same
for the metrics on both sides of the hypersurface.
Parametrizing the hypersurface
with the coordinates $y^a=(t,\vartheta,\varphi)$,
the hypersurface metric tensor $h_{ab}$
as induced from the metric (\ref{eq:gends2})
is given by $h_{ab}=g_{\alpha\beta} \, e^{\alpha}_a e^{\beta}_b$,
where $e^{\alpha}_a = \partial x^{\alpha} / \partial y^a$.
The hypersurface line element is given by
  \begin{equation}
  \rmd s^2 = h_{ab} \, \rmd y^a \, \rmd y^b
          = - k \, \big( 1 - \mu(a) \big) \, \rmd t^2
            + a^2 \, \rmd\Omega^2.
  \end{equation}
Therefore, for the first junction condition to hold we must have
$ k_{\mathrm{int}}(1-\mu_{\mathrm{int}}(a)) =
k_{\mathrm{ext}}(1-\mu_{\mathrm{ext}}(a)) $.
This is accomplished by setting $k_{\mathrm{ext}}=1$
for compatibility with the usual form
of the exterior Schwarzschild metric in case $\lambda_{\mathrm{ext}}=0$,
and setting $k_{\mathrm{int}}=(1-y)/(1-x)$
to appropriately rescale the time coordinate in the interior.
The second junction condition requires that for the smooth joining
of the metrics on the two sides of the hypersurface
the extrinsic curvature on the hypersurface,
$K_{ab} = n_{\alpha;\beta} e^{\alpha}_a e^{\beta}_b$,
be the same on both sides of the hypersurface.
The nonvanishing components of the extrinsic curvature
for the hypersurface of radius $r=a$
embedded in the metric of the form (\ref{eq:gends2}) are
  \begin{equation}
  K^{t}_{t} = - \frac{ \mu'(a) }{ 2 \, \sqrt{1-\mu(a)} }
  \qquad \mathrm{and} \qquad
  K^{\vartheta}_{\vartheta} = K^{\varphi}_{\varphi} = \frac1a \sqrt{1-\mu(a)}.
  \end{equation}
In the gravastar the second junction condition is not satisfied.
The joining of the metrics is not smooth, but is still possible
if one allows for the energy-momentum distribution on the hypersurface.
The standard expression for the shell energy-momentum tensor
on a timelike hypersurface is
  \begin{equation}
  S^{a}_{b} = - \frac1{8\pi} \left(
    \big[ K^{a}_{b} \big] - \big[ K \big] \delta^a_b \right),
  \end{equation}
where the brackets denote the discontinuity of the quantity
over the hypersurface and $K=K^a_a$.
Using the compactness functions for the exterior and interior metric
given by (\ref{eq:muint}) and (\ref{eq:muext}) and the notation (\ref{eq:xyz}),
the surface energy density can be written
  \begin{equation}
  \sigma = - S^{t}_{t} = - \frac{1}{4\pi\,a} \left[ \sqrt{1-\mu(a)} \right]
                       = \frac{1}{4\pi\,a} \,
                         \left( \sqrt{1-x} - \sqrt{1-y} \right),
  \label{eq:sigma}
  \end{equation}
and the isotropic surface tension can be written
  \begin{eqnarray}
  \theta = - S^{\vartheta}_{\vartheta} = - S^{\varphi}_{\varphi}
       & = - \frac{1}{8\pi\,a} \left[
         \frac{1-\mu(a)-r\mu'(a)/2}{\sqrt{1-\mu(a)}} \right] \nonumber \\
       & = \frac{1}{8\pi\,a} \, \left( \frac{1-2x}{\sqrt{1-x}}
           - \frac{1-\frac12y-\frac32z}{\sqrt{1-y}} \right).
  \label{eq:theta}
  \end{eqnarray}
Note that by definition the surface tension
has the opposite sign of the surface pressure,
i.e., $\theta>0$ indicates that the surface (shell) is `under tension',
while $\theta<0$ indicates that it is `under pressure'.

\section{Energy conditions \label{sec:ec}}

The four standard energy conditions of GR formulated in terms of
the components of the energy-momentum tensor of the gravastar shell
--- the surface energy density $\sigma$ (\ref{eq:sigma})
and isotropic surface tension $\theta$ (\ref{eq:theta}) ---
read as follows
  \begin{itemize}
  \item Weak energy condition (WEC):
        $\sigma \ge 0$ and $\sigma - \theta \ge 0$,
  \item Null energy condition (NEC):
        $\sigma - \theta \ge 0$,
  \item Strong energy condition (SEC):
        $\sigma - \theta \ge 0$ and $\sigma - 2 \theta \ge 0$,
  \item Dominant energy condition (DEC):
        $\sigma \ge 0$ and $\sigma - |\theta| \ge 0$.
  \end{itemize}
One can see from the structure of (\ref{eq:sigma}) and (\ref{eq:theta})
that the status of the energy conditions for the gravastar shell
can be studied in the space of the three configuration variables,
$x,y,z\in[0,1)$.
As pointed out in Introduction,
we require that the energy momentum tensor of the gravastar shell
satisfies all energy conditions except the SEC.
Since the DEC is more restrictive than the WEC or the NEC,
we proceed to consider the DEC through a number of steps.

We begin with the condition of non-negative surface energy density which,
as can be easily seen from (\ref{eq:sigma}), reduces to
  \begin{equation}
  \sigma \ge 0 \quad \Longrightarrow \quad x\le y.
  \label{eq:nnsigma}
  \end{equation}
This requirement has already been encoded in (\ref{eq:zxy}).

Continuing with the condition $\sigma - |\theta| \ge 0$,
we break it into cases according to $\theta\lesseqgtr0$.
We find
  \begin{equation}
  \theta \lesseqgtr 0 \quad \Longrightarrow \quad x \gtreqless f_0(y,z),
  \label{eq:cond:theta}
  \end{equation}
where the function $f_0$ is given by
  \begin{eqnarray}
  f_0(y,z) = \frac1{32} \Big( 13 + y + 6z - \frac{(1-3z)^2}{1-y} \nonumber \\
  \mbox{} - \frac{2-y-3z}{1-y} \,
  \sqrt{ 36 + y^2 - 6y(6-z) - 3z (4-3z) } \Big).
  \label{eq:fzero}
  \end{eqnarray}
The configurations with $\theta=0$ (vanishing surface tension/pressure),
satisfying $x=f_0(y,z)$, are understood as the shells of dust.
The DEC is clearly satisfied for dust shells provided (\ref{eq:nnsigma}) holds.
For the case $\theta < 0$, i.e. shells under (positive) surface pressure,
the condition $\sigma - |\theta| \ge 0$ reduces to
  \begin{equation}
  \theta < 0, \quad \sigma + \theta \ge 0 \quad
  \Longrightarrow \quad f_0(y,z) < x \le f_+(y,z),
  \label{eq:cond:cew}
  \end{equation}
where $f_+$ is given by
  \begin{eqnarray}
  f_+(y,z) = \frac1{128} \Big( 61 + 25y + 30z
                               - \frac{(1-3z)^2}{1-y} \nonumber \\
  \mbox{} - \frac{6-5y-3z}{1-y} \,
  \sqrt{ 100 + 25y^2 - 9(4-z)z - 2y (62-15z) } \Big).
  \label{eq:fplus}
  \end{eqnarray}
For the case $\theta>0$, i.e.\ shell under (positive) tension,
$\sigma - |\theta| \ge 0$ reduces to
  \begin{equation}
  \theta>0, \quad \sigma-\theta \ge 0 \quad
  \Longrightarrow \quad f_-(y,z) \le x < f_0(y,z),
  \label{eq:cond:wec}
  \end{equation}
where
  \begin{equation}
  f_-(y,z) = 1 - \frac{4(1-y)}{(2-3y+3z)^2}.
  \label{eq:fminus}
  \end{equation}
Combining the above cases with the condition (\ref{eq:nnsigma})
allows us to summarize the DEC (which includes the WEC and the NEC)
with the following conditions on the configuration variables $x,y,z\in[0,1)$:
  \begin{equation}
  x \le y, \qquad
  \cases{ f_0(y,z) < x \le f_+(y,z) & $\theta < 0$ \quad (pressure) \\
          x = f_0(y,z) & $\theta = 0$ \quad (dust)\\
          f_-(y,z) \le x < f_0(y,z) & $\theta > 0$ \quad (tension)\\}
  \label{eq:dec}
  \end{equation}
The configurations saturating the DEC with $x=f_+$ and $x=f_-$
can be called stiff and anti-stiff shells.
The implications of these conditions on the allowed
gravastar configurations are studied in the next Section.

\section{Allowed configurations \label{sec:allowed}}

The shape of the functions $f_0$, $f_+$ and $f_-$,
given by (\ref{eq:fzero}), (\ref{eq:fplus}) and (\ref{eq:fminus}),
determines the regions in the space of configuration variables
$x,y,z\in[0,1)$ that are allowed by the DEC (\ref{eq:dec}).
In addition, we recall the `gravastar condition' $z\le x$,
expressing the requirement that the (dark) energy density inside the gravastar
is not less than the energy density in the exterior spacetime.
Before dividing the analysis into cases
corresponding to specific ranges of the configuration variable $z$,
we observe some general properties of $f_0$, $f_+$ and $f_-$.

As already pointed out in Sec.~\ref{sec:model},
$y<z$ implies $M<0$ (see Eq.~(\ref{eq:muext})),
but it can also be shown that
  \begin{equation}
  f_-(y,z) > f_+(y,z), \quad 0 \le y < z < 1,
  \end{equation}
i.e., the DEC (\ref{eq:dec}) cannot be satisfied with $y<z$.

For $y=z$, which corresponds to $M=0$, one can show that
  \begin{equation}
  f_-(y,z) = f_0(y,z) = f_+(y,z) = z, \qquad 0 \le z  = y < 1,
  \label{eq:gen2}
  \end{equation}
indicating that the only $y=z$ configuration satisfying the DEC is $x=y=z$,
which further implies $\lambda_{\mathrm{int}} = \lambda_{\mathrm{ext}}$,
i.e., simply de~Sitter spacetime with no gravastar.

For $y>z$ it can be shown that the following relations hold,
  \begin{equation}
  f_0(y,z) < f_+(y,z) < y, \qquad 0<z<y<1.
  \label{eq:gen3}
  \end{equation}
This indicates that the DEC satisfying configurations with $\theta<0$
(shell under surface pressure) exist for all $z$.

In Fig.~\ref{fig:one} the DEC satisfying gravastar configurations
are indicated with the shaded regions in the $x,y$ plane
for several fixed values of $z$.
Details are discussed in what follows.

\subsection{ $z=0$ }

With $z=0$ we are actually considering the gravastar
with the Schwarzschild exterior geometry.
One can show that $f_0(y,0)<0$ for $0<y<1$,
meaning that in this case the dust shells
and the shells under surface tension
cannot satisfy the condition (\ref{eq:zxy}).
Inspection of $f_+(y,0)$ reveals that it is greater than zero
for $0<y<24/25$ with a maximum at $y=4/5$,
  \begin{equation}
  f_+(4/5,0) = \frac{1}{32}\big(19-\sqrt{105}\big) \simeq 0.274.
  \label{eq:z0xmax}
  \end{equation}
It follows that the $z=0$ gravastar configurations allowed by the DEC
involve the shell under surface pressure and are constrained by
  \begin{equation}
  0 = z \le y \le 24/25, \qquad 0 \le x \le f_+(y,0).
  \end{equation}
The corresponding region in the $x,y$ plane is shown
in Fig.~\ref{fig:one} (upper-left plot).

In the original Mazur--Mottola gravastar model it is assumed
that the compactness on both sides of the vacuum phase transition layer
will approach unity (horizon formation) to induce the vacuum phase transition,
i.e.~$x,y\to 1$.
As we can see from the above analysis,
the DEC satisfying configurations are
clearly pushed away from the Mazur--Mottola limit.
The highest compactness on the exterior side of the gravastar shell
that can be reached without violating the DEC is $y_{\max}=24/25$.
It corresponds to the surface gravitational redshift
$ Z_{\max} = (1-y_{\max})^{-1/2} - 1 = 4 $,
and is obtained with flat interior geometry.
The same upper bound on the surface compactness was obtained
in a similar context \cite{FrauHoenKon90} where the DEC-satisfying shell
was considered outside of a Schwarzschild black hole of mass $m$,
in the limit where $m\to0$.
The finding which is relevant in the context of the gravastar model
is that the introduction of the de~Sitter geometry
(dark energy) in the interior,
relative to the flat interior geometry (hollow shell),
does not help to support the DEC-satisfying shell against the collapse
at high values of the surface compactness,
as it might be naively expected
due to the repulsive character of the de~Sitter interior
(free falling particle accelerating away from the centre).
We have also obtained the upper bound on the compactness
at the interior side of the shell;
the highest $x$ that can be reached without violating the DEC
is $x_{\max} \simeq 0.274$, given by (\ref{eq:z0xmax}),
lowering the maximal exterior compactness to $y=4/5$
(surface redshift $Z=\sqrt5-1\simeq1.236$).
Let us also note that in the model of Mazur and Mottola 
the vacuum phase transition shell is contributing
a negligible fraction to the net gravitational mass of the gravastar.
Here we see that such a condition cannot be met
if the DEC is to be satisfied.

\subsection{ $0<z<1/3$ }

With $z>0$ we are introducing
the Schwarzschild--de~Sitter exterior geometry.
Let us begin with the configurations with the gravastar shell
under (positive) surface pressure ($\theta<0$).
The functions $f_0$ and $f_+$ tend to $-\infty$ as $y\to1$,
but they have a maximum in $y$ with the value greater than $z$ for $z<y<1$.
The area in the $x,y$ plane bounded by $f_0$ and $f_+$
is the $\theta<0$ DEC satisfying region,
but we must also observe the `gravastar requirement' $x \ge z$.
Therefore we look for the intersections
of $f_0$ and $f_+$ with the $x=z$ line and we find
  \begin{equation}
  f_0(y_0,z) = z, \qquad y_0 = y_0(z) = \frac{z\,(5-9z)}{1-z}
  \label{eq:2y0}
  \end{equation}
and
  \begin{equation}
  f_+(y_+,z) = z, \qquad y_+ = y_+(z) = \frac{24-19z-9z^2}{25\,(1-z)}.
  \label{eq:2y+}
  \end{equation}
The DEC satisfying gravastar configurations with the shell under
surface pressure ($\theta<0$) are therefore
  \begin{equation}
  0 < z < 1/3, \quad
  \cases{ f_0 < x \le f_+(y,z) & for $z < y \le y_0(z)$ \\
          z \le x \le f_+(y,z) & for $y_0 < y \le y_+$(z) \\ }
  \end{equation}
where $y_0$ and $y_+$ are given by (\ref{eq:2y0}) and (\ref{eq:2y+}).
The dust shells ($\theta=0$) configurations are those with
  \begin{equation}
  0<z<1/3, \quad x = f_0(y,z), \quad z < y \le y_0.
  \end{equation}
Upon showing that $f_-(y,z)<z$ for $z<y<y_0(z)$
it follows that the DEC satisfying configurations
with the shell under surface tension ($\theta>0$) are
  \begin{equation}
  0<z<1/3, \quad z < x < f_0(y,z), \quad z \le y \le y_0.
  \end{equation}
The DEC satisfying gravastar configurations for the case $z=1/4$
are shown in Fig.~\ref{fig:one} (upper-right plot).

Considering $z>0$ puts the gravastar model into the context
where positive cosmological constant $\Lambda$
is present in the Einstein equations.
Here we can no longer discuss the surface gravitational redshift,
but we still find it important to discuss the maximal surface compactness
that can be reached without violating the DEC.
As in the $z=0$ case, the highest compactness on the
exterior side of the shell is achieved with the `hollow interior', $x=z$,
which here means that the (dark) energy density in the interior
is equal to that in the exterior;
the maximal compactness is $y_+(z)$ given in (\ref{eq:2y+}).
It can be shown that $y_+(z)$ starts with the value $24/25$ at $z=0$
and increases toward unity as $z\to1/3$.
However, if the (dark) energy density in the interior
is higher than that in the exterior region, $x>y$,
the maximal compactness that can be achieved
is lower than that given by $y_+$.
Another interesting feature of gravastar configurations with $z>0$ is that
in addition to the configurations with shell under surface pressure
here we can have shells of dust and shells under surface tension.

\subsection{ $z=1/3$ \label{sec:43} }

The $z=1/3$ case can be seen as the critical case
between the regimes $z\lessgtr1/3$.
The functions $f_0(y,1/3)$ and $f_+(y,1/3)$,
bounding the $\theta<0$ DEC satisfying region,
are monotonically increasing in $y$ for $1/3<y<1$,
reaching finite values $f_0(1,1/3)=1/2$ and $f_+(1,1/3)=3/4$,
while $f_-(y,1/3)$ is monotonically decreasing
diverging to $-\infty$ as $y\to1$.
Observing also the `gravastar requirement' $x\ge z$,
the DEC satisfying configurations are simply
  \begin{equation}
  1/3 = z \le y < 1, \qquad
  \cases{ f_0(y,z) < x \le f_+(y,z) & ($\theta < 0$) \\
          x = f_0(y,z) & ($\theta = 0$) \\
          z \le x < f_0(y,z) & ($\theta > 0$) \\}
  \end{equation}
and are shown in Fig.~\ref{fig:one} (lower-left plot).
The distinctive feature of this special case $z=1/3$
is that there is a wide range of configurations
that allow the surface compactness on the exterior side
to approach unity arbitrarily close.
The compactness on the interior side of the shell is still
bounded from above by $f_+$, or fixed by $f_0$ for a dust shell.
For example, the dust shell with $y\to1$ will have $x\to1/2$,
which implies interior (dark) energy density
which higher by factor $x/z\to3/2$
than the (dark) energy density in the exterior.
However, the value of $z=1/3$ considered here
implies that the radius of the gravastar is larger
than one half of the radius of the exterior horizon
of the Schwarzschild--de~Sitter geometry,
which makes these configurations highly unlikely
to be relevant in the astrophysical context.

\subsection{ $1/3<z<1$ }

\begin{figure}
\includegraphics[scale=1.0]{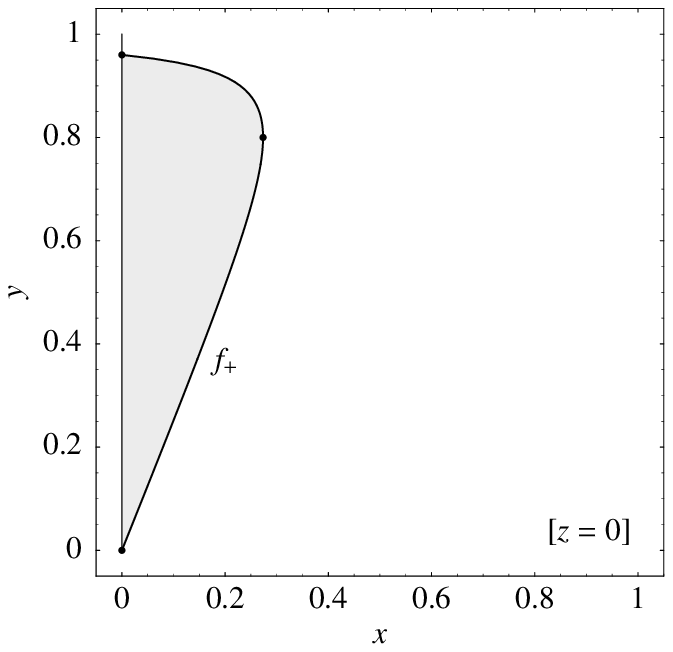}
\includegraphics[scale=1.0]{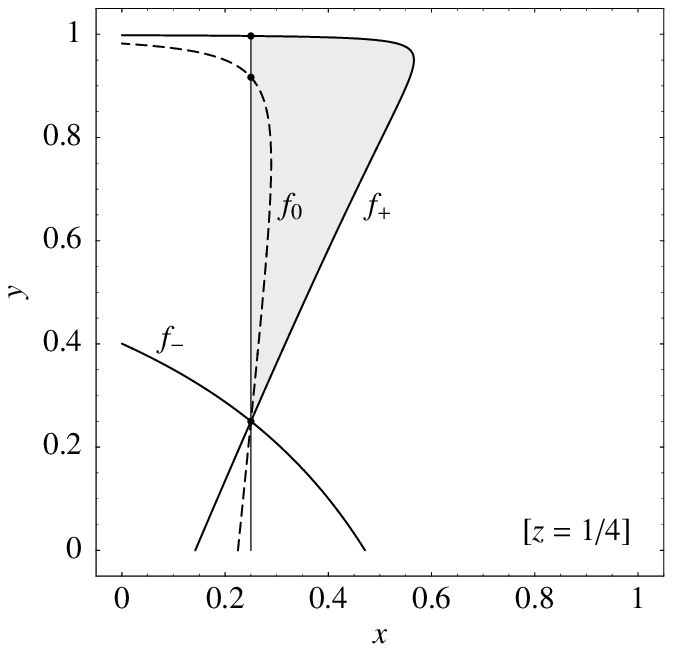}\\
\includegraphics[scale=1.0]{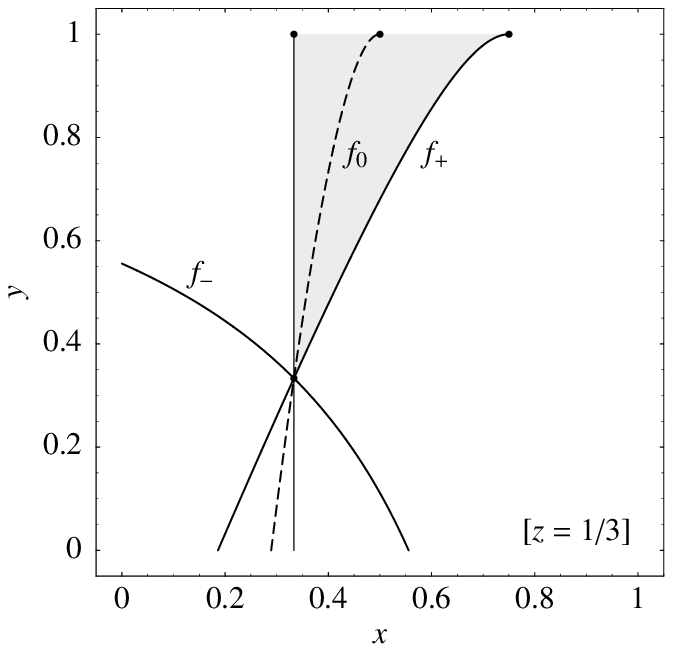}
\includegraphics[scale=1.0]{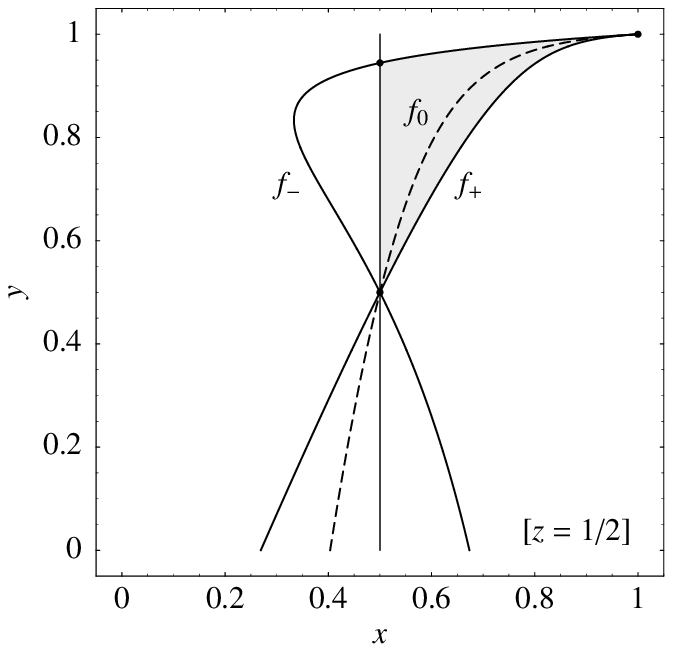}
\caption{\label{fig:one}
Gravastar configurations allowed by the DEC
shown as shaded regions in the $x,y$ plane 
for fixed values of $z=0,1/4,1/3,1/2$.
($x,y,z$ are defined in (\ref{eq:xyz}))}
\end{figure}

For completeness here we consider also the $1/3<z<1$ configurations.
The functions $f_-$, $f_0$ and $f_+$ tend to $1$ as $y\to 1$.
Since $f_0 \ge z$ for $1/3<y<1$,
the configurations with dust shells and
shells under surface pressure ($\theta\le0$) are
  \begin{equation}
  1/3 < z < y < 1, \qquad f_0(y,z) \le x \le f_+(y,z).
  \end{equation}
The shell under tension ($\theta>0$) configurations
must be treated more carefully since it can be shown
that the function $f_-(y,z)$ dips into the $x<z$ region for $1/3 < z < 2/3$,
which is forbidden by the `gravastar requirement' $z\le x$.
We find
  \begin{equation}
  f_-(y_-,z) = z, \qquad y_- = y_-(z) = \frac{8-3z-9z^2}{9(1-z)},
  \label{eq:4y-}
  \end{equation}
so for $1/3 < z < 2/3$, the $\theta>0$ DEC satisfying gravastar
configurations are
  \begin{equation}
  1/3 < z < 2/3, \quad
  \left\{
    \begin{array}{cc}
    z < y \le y_-, & z \le x < f_0(y,z)\\
    y_- < y < 1, & f_-(y,z) \le x < f_0(y,z)
    \end{array}
  \right.
  \end{equation}
where $y_-$ is given by (\ref{eq:4y-}).
For $2/3 \le z < 1$ the $\theta>0$ DEC satisfying gravastar
configurations are
  \begin{equation}
  2/3 \le z < y < 1, \qquad f_-(y,z) \le x < f_0(y,z).
  \end{equation}
The special case $z=1/2$ is shown in Fig.~\ref{fig:one} (lower-right plot).
The interesting feature of $z>1/3$ configurations
is that the compactness on both sides of the shell
can approach unity without violating the DEC.
However, as already commented in the case of $z=1/3$ configurations,
the size of such object is too large to represent
a body of astrophysical interest.

\section{A note on the equation of state \label{sec:eos}}

In a perfect fluid described by an equation of state of the form $p=p(\rho)$,
where $p$ is the isotropic pressure and $\rho$ is the energy density,
the speed of sound propagation, $c_{\mathrm{s}}$,
is given by the relation $c_{\mathrm{s}}^2=\rmd p / \rmd \rho$.
In order for the causality to be preserved,
it is natural to require that the sound speed
does not exceed the speed of light, i.e.\ in our units $c_{\mathrm{s}}<1$.
For a view on the status of the relation among causality
and the speed of sound see Ref.\ \cite{EllisCausality07}.
Here we will investigate the speed of sound
propagating (tangentially) along the gravastar shell,
assuming that the equation of state for matter comprising the shell
can be given in the simple (barotropic) form $\theta=\theta(\sigma)$.
Note that while the matter comprising the shell
can be considered as the perfect fluid on the hypersurface,
in the complete spacetime it represents a highly anisotropic structure
(the transverse pressure is described by the $\delta$-distribution).

In Ref.~\cite{VissGS1} a procedure has been developed
that allows extracting of the equation of state $\theta=\theta(\sigma)$
of the matter comprising the gravastar shell
on the basis of the requirement
that the shell be stable against radial perturbations.
From the dynamical version of the expressions
for the surface energy density and the surface tension of the shell,
which involve the terms $\dot{a}^2$ and $\ddot{a}$,
overdot indicating the derivative with respect to the proper time
of the observer co-moving with the shell,
the `classical' equation of motion of the form $\frac12 \dot{a}^2+V=E$ follows.
The `potential' $V$ is a function of the gravastar radius $a$
and other configuration variables, while the `energy' $E\equiv0$
(one is not allowed to adjust the energy as in the classical context).
Upon identifying $\dot{a}^2=-2V(a)$ and $\ddot{a}=-V'(a)$
(See Ref.~\cite{VissGS1} for the full derivation)
the dynamical expressions for $\sigma$ and $\theta$
become functions of the dynamical gravastar radius.
The requirement for the stability of a shell at the radius $a$
can then be expressed as the requirement imposed on the potential,
  \begin{equation}
  V(a) = 0, \qquad V'(a) = 0, \qquad V''(a) \ge 0.
  \end{equation}
Therefore, if one chooses a particular shape of the potential $V(a)$
leading to stable gravastar shells,
the equation of state $\theta=\theta(\sigma)$ can be parametrically
extracted by varying $a$ over the range of interest.
The important special case is $V(a)\equiv0$ which,
as argued in Ref.~\cite{VissGS1},
reflects the notion of strict stability of the shell.
If the shell would be displaced to a nearby radius,
it would find itself there in a new state of equilibrium.
In the case $V(a)\equiv0$ the dynamical equations for surface energy density
and surface pressure of the shell coincide with (reduce back to)
the static expressions for $\sigma$ and $\theta$
given by (\ref{eq:sigma}) and (\ref{eq:theta}).
Therefore we can use this approach to compute the quantity
  \begin{equation}
  c_{\mathrm{s}}^2 = - \frac{\rmd \theta}{\rmd \sigma}
  = - \frac{\rmd \theta / \rmd a}{\rmd \sigma / \rmd a}
  \label{eq:cs2def}
  \end{equation}
directly from (\ref{eq:sigma}) and (\ref{eq:theta}).

\begin{figure}
\includegraphics[scale=1.0]{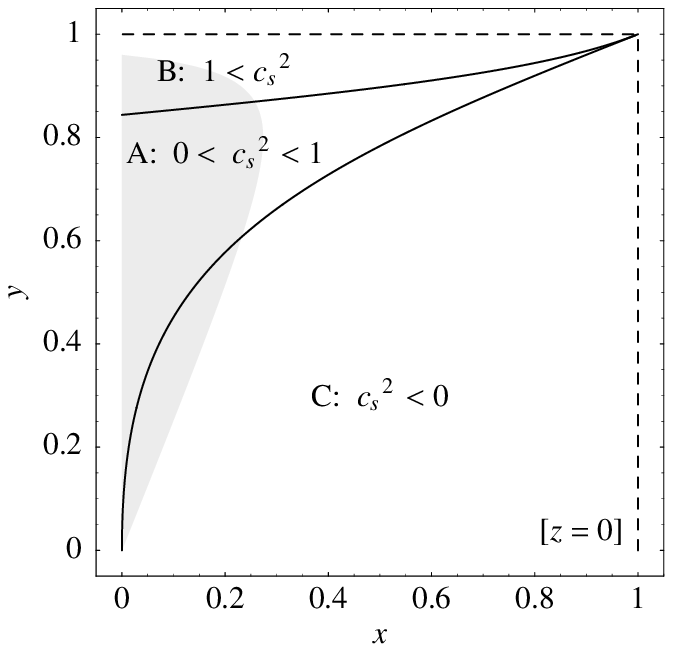}
\includegraphics[scale=1.0]{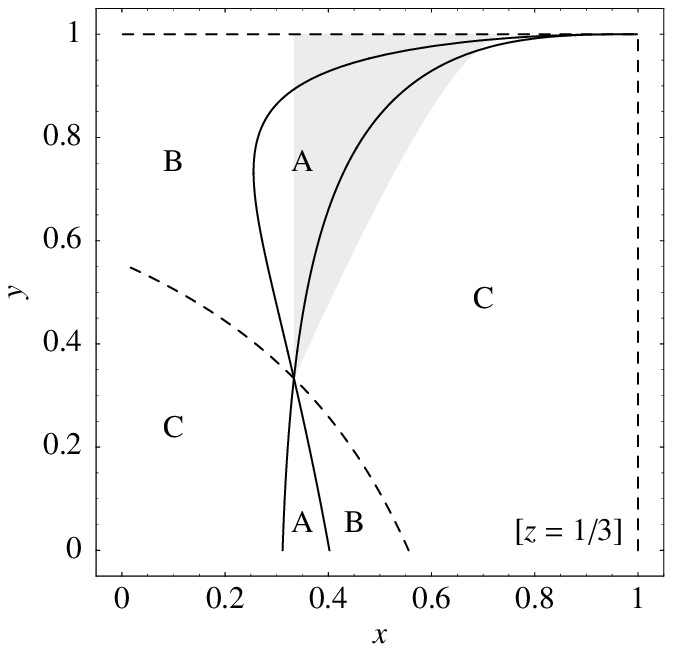}
\caption{\label{fig:two}
Behaviour of the quantity $c_{\mathrm{s}}^2=-\rmd\theta/\rmd\sigma$
(speed of sound squared) as derived from the requirement for the
stability against the radial perturbations
for the gravastar shells with $z=0$ (left plot)
and $z=1/3$ (right plot), shown in the $x,y$ plane.
Dashed lines indicate configurations where $c_{\mathrm{s}}^2\to\pm\infty$,
while the solid lines bound the region `A'
in which $0<c_{\mathrm{s}}^2<1$.
Region `B' is the $1 < c_{\mathrm{s}}^2$ superluminal regime,
while `C' is the $c_{\mathrm{s}}^2<0$ unstable regime.
Shaded regions are the gravastar configurations allowed by the DEC
(See Fig.~\ref{fig:one}).}
\end{figure}

We now proceed to compute $c_{\mathrm{s}}^2$ for the gravastar models
we have considered so far.
Using (\ref{eq:cs2def}), (\ref{eq:sigma}) and (\ref{eq:theta})
one can obtain an expression for $c_{\mathrm{s}}^2$
in terms of $a$, $\lambda_{\mathrm{int}}$, $\lambda_{\mathrm{ext}}$ and $M$
which is rather involved,
but with the use of our configuration variables $x,y,z$
defined in (\ref{eq:xyz}) it becomes much more compact and can be written
  \begin{equation}
  c_{\mathrm{s}}^2 = \frac{
    4 v^3 - u^3 [ 1 - 6 z + 3 ( v^2 + z ) ( v^2 + 3 z ) ]
  }{ 4 u^2 v^2 [ - 2 v + u ( - 1 + 3 v^2 + 3 z ) ] }
  \label{eq:cs2uvz}
  \end{equation}
where $u^2 = 1 - x$ and $v^2 = 1 - y$.
The quantity $c_{\mathrm{s}}^2$ diverges for $u\to0$ and $v\to0$
and also at $u = 2 v / (3 v^2 + 3 z - 1)$.
To obtain the boundaries of the regions in the $x,y$ plane
(for the given values of $z$) in which $0\le c_{\mathrm{s}}^2 \le 1$
one must solve $c_{\mathrm{s}}^2=0,1$ for $u$ which leads to a cubic equation.
The resulting expressions are very complicated
so we only plot the results for $z=0$ (Schwarzschild exterior geometry)
and $z=1/3$ (case considered in Sec.~\ref{sec:43}) in Fig.~\ref{fig:two}.
The region where causality is preserved
with the equation of state having usual properties
(surface pressure increases as energy density increases)
are indicated by `A'.
We can see that the causality requirement
combined with the energy conditions considered before
imposes still stronger upper bounds
on the gravastar exterior surface compactness $y$.
As an example, for the $x=z=0$ case (hollow shell, Schwarzschild exterior)
the highest allowed surface compactness is $ y_{\max} \simeq 0.848 $
or analytically $2(106 + \alpha - 89/\alpha)/255$
where $\alpha=(225\sqrt{106}-2195)^{1/3}$.
In both examples shown in Fig.~\ref{fig:two}
we also see that the region `C' where $c_{\mathrm{s}}^2$ is negative
includes some regions that are allowed by the energy conditions.
Negative $c_{\mathrm{s}}^2$ means that the surface pressure decreases
as the surface energy density increases,
which indicates instability with respect to perturbations.
Therefore, the requirement that the dependence of $\theta$ on $\sigma$
is such that stability with respect to radial perturbations is obtained
can lead to configurations where the gravastar shell is unstable against
tangential perturbations such as sound propagation.

\section{Conclusions \label{sec:conc}}

Compact objects like gravastars
envisaged as possible alternatives to black holes
are certainly not less controversial
than black holes themselves \cite{Lasota06}.
The essential phenomenological feature of a gravastar
that makes it difficult to distinguish such a body
from a true black hole is its high surface redshift.
Also, the analysis of radiation from a supermassive object Sgr~A$^*$
\cite{BroderickNarayan05}
implied the existence of a horizon in this object.
However, the absence of detectable heating
could also be consistent with a gravastar
possesing sufficiently large heat capacity
\cite{BroderickNarayan07}.
Another feature which may help in this respect
is the analysis of quasi-normal modes of oscillation of a gravastar.
They are found to be different
from those of a black hole \cite{ChirentiRezzolla07},
and could, in principle, provide a way
to observationally discern a gravastar from a black hole.
In the same spirit it is important to analyze
the constraints on possible realisations of gravastar models.
Since the physical theory behind the quantum vacuum phase transitions
which are assumed to take place within the gravastar shell
has not been fully formulated,
the energy conditions of GR provide the most general
model-independent constraint on the structure
of the energy-momentum tensor of the gravastar.
We have carried out a detailed analysis of the constraints
that the energy conditions impose on the possible configurations
of the single $\delta$-shell version of the gravastar model \cite{VissGS1}.

The most relevant constraint is the upper bound
on the surface compactness $\mu=2m/r$ of the gravastar
which was in the original MM model assumed to approach unity arbitrarily close.
Exactly this feature gives raise
to arbitrarily large surface gravitational redshift
which makes the gravastar observationally difficult to distinguish
from the true Schwarzschild black hole.
(Note that the original MM model violates the DEC
at its two $\delta$-shells that are assumed to possess
nonvanishing surface pressure but vanishing surface energy density.
For a general discussion of known violations of energy conditions
see eg.\ Refs~\cite{VisserBook,BarceloVisser02}).

In this paper we have derived the bounds on the surface compactness
that follow from the requirement
that the matter comprising the gravastar shell satisfies the DEC.
The procedure is fully analytic and has been made possible
by the use of the configuration variables $x,y,z$
based the values of the compactness at the inner and the outer
side of the shell defined in (\ref{eq:xyz}).
The space of gravastar configurations satisfying the DEC
is confined within the range $f_-(y,z) \le x \le f_+(y,z)$,
where the functions $f_-$ and $f_+$ identifying
the configurations with anti-stiff and stiff shells
are given by (\ref{eq:fminus}) and (\ref{eq:fplus}).
In case of the Schwarzschild exterior geometry
the maximal surface compactness $\mu_{\max}=y_{\max}=24/25$
is achieved with a stiff shell and flat geometry in the gravastar interior
(compactness in the interior side of the shell $x=0$),
i.e.\ in the limit of vanishing (dark) energy density in the interior.
Introducing the (dark) energy density in the interior ($x>0$)
does not lead to higher surface compactness that can be reached.
This finding may be seen as counter-intuitive
since one might have expected that the repulsive character of the
de~Sitter geometry will help to support the shell against the collapse.
however, as our analysis has shown,
this is not the case if the shell is to satisfy the DEC.

For completeness, we have extended our analysis
to the case of the Schwarzschild--de~Sitter exterior geometry
which places the gravastar model in the context
where the positive cosmological constant $\Lambda$
is present in the Einstein equations.
Relevant bounds on the surface compactness
have been obtained for all values of the configuration variable
$z=\Lambda a^2 / 3$, $a$ being the gravastar radius.
For $z<1/3$ the highest surface compactness
is bounded below unity and is achieved with the stiff shell.
At $z=1/3$ there is a wide range of configurations 
where the surface compactness can approach unity arbitrarily close
and it includes shells under surface pressure, dust shells
and shells under surface tension.
For $z>1/3$ the highest surface compactness is again bounded below unity
and is achieved with anti-stiff shells.

We have also applied the procedure of Ref.~\cite{VissGS1}
to compute the quantity $c_{\mathrm{s}}^2=-\rmd \theta / \rmd \sigma$
which in the context of perfect fluids corresponds to the speed of sound.
In order not to violate the principle of causality
one normally requires that the sound speed does
not exceed the speed of light.
This requirement has given still stronger upper bounds
on the gravastar surface compactness
as compared to the bounds derived from the energy conditions.
However, we must emphasize that the procedure is based on the assumption
that the exotic matter comprising the gravastar shell
obeys a barotropic equation of state of the form $\theta=\theta(\sigma)$,
on the assumption that the shell is stable against the radial perturbations,
and most importantly on the assumption
that the fluctuations in surface tension and surface energy density
propagate along the shell at the speed given by the usual formula stated above.
Since the full theory describing the gravastar shell is not available
none of these assumptions can be taken for granted.
Similar situation has been encountered in the context of the
stability analysis of wormhole solutions in Ref.~\cite{PoissonVisserWH95}.

All the above results were derived using the $\delta$-shell gravastar model,
but they may be relevant in the broader context of the dark energy stars,
or gravastars with continuous pressure profiles,
where one attempts to construct astrophysically plausible models
of self-gravitating objects with dark-energy cores.
As shown in Ref.~\cite{VissGS2}, and applied in in Ref.~\cite{DeBenedGS1},
such objects require anisotropic principal pressures.
The analysis of the maximal surface compactness of anisotropic spheres
is considerably more complicated than in the isotropic case;
probably the most general results obtained in this area is Ref.~\cite{Hakan07},
although still not general enough
to include the case of the dark energy interior,
or Ref.~\cite{BoeHar06b} where the influence of the cosmological constant
on the compactness and other physical parameters
of anisotropic configurations was considered.
However, thin shells (soap bubbles) appear to be the
most efficient configurations in attempts to reach high surface compactness
of anisotropic spherical bodies.
The maximal compactness of the `shell around a black hole'
of Ref.~\cite{FrauHoenKon90},
or that of our flat interior gravastar (hollow shell),
clearly saturates the general bound derived in Ref.~\cite{Hakan07},
and similar suggestions have also been given in Ref.~\cite{Bondi99}.

\section*{Acknowledgments}

We acknowledge the support from the Croatian Ministry of Science
under the project 036-0982930-3144.
SI thanks for hospitality the University of Vienna
where part of this work was carried out.

\section*{References}

\providecommand{\newblock}{}

\end{document}